\documentclass[12pt]{iopart}
\usepackage{iopams} 
\usepackage[dvips]{graphicx}
\usepackage[usenames,dvipsnames]{color}
\usepackage{multirow}

\begin{document}
\title{Prediction of first-order martensitic transitions in strained epitaxial films}

\author{S Sch\"onecker$^{123}$, M Richter$^{14}$, K Koepernik$^1$ and H Eschrig$^{1\dagger}$}
\address{$^1$IFW Dresden, P.O.\,Box 270116, D-01171 Dresden, Germany}
\address{$^2$KTH Royal Institute of Technology, Brinellv\"agen 23, SE-10044 Stockholm, Sweden}
\ead{$^3$stesch@kth.se}
\ead{$^4$m.richter@ifw-dresden.de}
\address{$^\dagger$deceased}

\begin{abstract}
Coherent epitaxial growth allows to produce strained crystalline films
with structures which are unstable in the bulk.
Thereby, the overlayer lattice parameters in the interface plane, $(a,b)$,
determine the minimum-energy out-of-plane lattice parameter, $c_{\rm min}(a,b)$.
We show by means of density-functional total energy calculations that
this dependence can be discontinuous and predict related
first-order phase transitions in strained tetragonal films of the elements
V, Nb, Ru, La, Os, and Ir.
The abrupt change of $c_{\rm min}$ can be exploited to switch
properties specific to the overlayer material.
This is demonstrated for the example of the superconducting critical
temperature of a vanadium film which we predict to jump by 20\% at a
discontinuity of $c_{\rm min}$.
\end{abstract}


\maketitle

\section{\label{sec:intro}Introduction}

Epitaxy is an important concept for the 
fabrication of films
with good crystalline quality like overlayers, multilayers,
compound materials, and ordered alloys. Such films are technologically 
important materials with adjustable electronic, magnetic, and
optical properties~\cite{Ohring:1992}. Epitaxial growth also allows to study fundamental
aspects of low-dimensional structures and interfacial effects, and it allows to produce
structures that are unstable in the bulk~\cite{Wuttig:2004}.
For example,
body-centered cubic (BCC) Co does not exist as a bulk phase but was stabilized
by epitaxial growth on a GaAs substrate~\cite{Prinz:1985}.

In coherent or pseudomorphic film growth, the in-plane film lattice parameters
are determined by the substrate-in-plane lattice parameters.
This allows to expose a material to static, non-isotropic but
homogeneous strain which is a valuable mean to
influence its intrinsic properties~\cite{Burkert:2004a}.
Experiments on ferromagnetic bulk-like films showed,
that the magnetic moments, the Curie temperature,
and the magnetic anisotropy can be tuned in a quasi-continuous
manner by varying the substrate~\cite{Buschbeck:2009,Kauffmann-Weiss:2011}.
Using piezoelectric substrates, even a continuous variation
of the lattice parameters and of the related film properties
was achieved~\cite{Doerr:2009,Rata:2008}.

Here, we predict by density-functional calculations
scanning 30 non-magnetic metallic elements,
that bulk-like epitaxial films of at least six
metals show first-order martensitic transitions
upon variation of the epitaxial strain.
For example, we find jumps in the lattice parameter of La by 
$0.83$ \AA{} and in the superconducting critical temperature of V by 20\%.

The examples considered in the present work include only the
simplest case of tetragonal non-magnetic monoatomic films.
We however emphasize that our main result, the possibility of isostructural
phase transitions~\cite{Rosner:2006,Zarechnaya:2010}
under epitaxial strain and a related
discontinuous dependence of intrinsic film properties on the interface
lattice parameters, is not restricted to certain structures, compositions,
or ground states.

\section{\label{sec:method}Method}

We model the structures of the considered films by
the epitaxial Bain path (EBP)~\cite{Alippi:1997}.
The EBP model applies to the bulk part (i.e., the interior) of a coherently
grown thick film~\cite{Ozolins:1998}.
It assumes that the film can be described by a bulk structure
with periodicity in all three dimensions.
Due to epitaxial coherency, the in-plane film lattice parameters
are rigidly tied to the substrate lattice parameters
in the substrate-film interface via the underlying
epitaxial relationship. On the other hand,
the film is free to relax its geometry perpendicular to the
interface due to the absence of forces.
Summarizing, the model neglects
all surface and interface effects except the determination of 
the in-plane film lattice parameters by the substrate.

In the past, the EBP model was used to predict metastable states
in transition
metals~\cite{Alippi:1997,Jona:2001,Jona:2002,Marcus:2002,Schoenecker:2011},
to investigate the magnetic order in strained
overlayers~\cite{Qiu:2001,Qiu:2001:E,Zeleny:2008,Tsetseris:2005}, and 
to identify the mother phases of strained bulk-like 
films~\cite{Alippi:1997,Ji:2003,Tian:1999,Marcus:1997}.
In the same context,
we recently suggested Ru, Os, and U films to be ferromagnetic
in certain ranges of epitaxial strain~\cite{Schoenecker:2012}.

\begin{figure}[tbp]
\begin{indented}
\item\centering\includegraphics[height=3cm,clip]{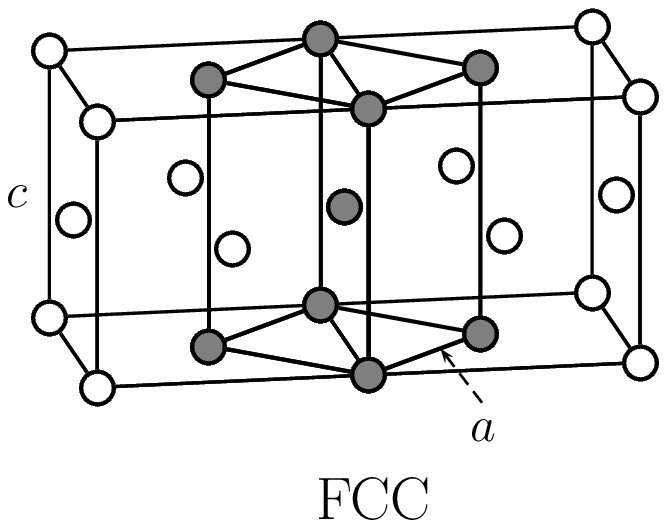}\qquad\qquad\includegraphics[height=3cm]{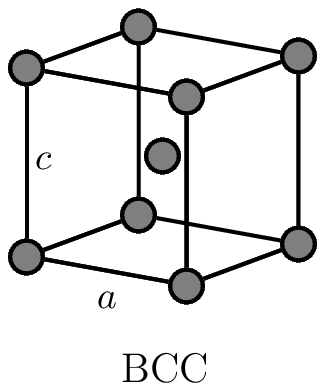}
\end{indented}
\caption{\label{fig:lattice}Delineation of the BCT lattice (closed symbols) in the FCC lattice ($c/a=\sqrt{2}$, open symbols, two unit cells shown) and in the BCC lattice ($c/a=1$). BCT lattice and BCC lattice are on top of each other.}
\end{figure}

Focusing on the important case of substrates with four-fold surface
symmetry and body-centered tetragonal (BCT) films with (001) orientation, 
we denote the in-plane film lattice parameter by $a$ and the out-of-plane
lattice parameter by $c$, see \fref{fig:lattice}.
The mentioned absence of forces perpendicular to the substrate-film
interface allows 
for a relaxation of $c$ minimizing the total film energy $E(a,c)$ 
at any fixed value of $a$~\cite{Alippi:1997,Ozolins:1998}.
Hence, the EBP of a certain BCT material is the set of points
$\{(a,c_{\rm min})\}\subset \{(a,c)\}$, for which
\begin{equation}
\textrm{EBP}\stackrel{\rm def}{=} \left\{ (a,c_{\rm min}) \Big| E\left(a,c_{\rm min}\right) = \min_c E(a,c) \right\}.
\label{eq:mydefEBP}
\end{equation}
The side condition determines the relaxation of $c$ for a given $a$ and 
defines the relation $c_{\rm min}(a)$.
Quantities traced \emph{along} the EBP only depend on $a$, e.g.,
the total energy $E_{\rm EBP}(a)\stackrel{\rm def}{=}E(a,c_{\rm min}(a))$.
The difference between $E_{\rm EBP}(a)$ and the ground state energy
of the film material,
$E_0$, corresponds to the general definition of an epitaxial strain energy~\cite{Ozolins:1998}.

A strict implementation of the notion 
`relaxation of $c$ minimizing $E(a,c)$'
leads to relevant, until now disregarded consequences for all materials
to which strained coherent epitaxy applies.
As we show, the function
 $E(a,c)|_a$ can have two equal-valued global minima at $c^1_{\rm min} \neq c^2_{\rm min}$
for certain values of $a = a_{\rm crit}$.
Thus, $c_{\rm min}(a)$ can be discontinuous at $a_{\rm crit}$ in a 
first-order transition.
This includes,
by epitaxial coherency, also the dependence of $c_{\rm min}$ on the
substrate lattice parameters in the interface plane.
In this way, the possibility to tune overlayer-material
specific properties can be restricted to a discontinuous manner.
However, one can
switch these properties on purpose,
if one succeeds to control the lattice parameter $a$ in a narrow
interval around $a_{\rm crit}$, e.g., by a piezoelectric substrate.

In the following, we verify the existence of degenerate global 
minima in $E(a,c)|_a$ by high-precision density-functional theory
(DFT) calculations, carried out with the full-potential local-orbital scheme
FPLO-7.00-28~\cite{Koepernik:1999}.
Using the local-density approximation according to Ref.~\cite{Perdew:1992} (PW92)
and a scalar-relativistic mode for elements with atomic number $< 49$, a full-relativistic mode otherwise,
we scanned the EBP of 30 elements with the atomic numbers 20-23, 29, 30, 38-48,
56, 57, 71-80, and 92 in a wide range of parameters $a$~\cite{Schoenecker:2011}.
The convergence of numerical parameters and the basis set were carefully 
checked~\cite{Schoenecker:2011}.
In order to converge the total energy per atom at a level smaller than $0.3$ meV, linear-tetrahedron
integrations with Bl\"ochl corrections were performed on a $24 \times 24 \times 24$ mesh in the full Brillouin zone, apart from the elements Cu, Ru, Ir, and Pt, for which a denser $48 \times 48 \times 48$ mesh was required.
All detected degenerate minima in $E(a,c)|_a$ were cross-checked by additional calculations employing the Perdew-Burke-Ernzerhof (PBE96) parameterization of the exchange-correlation functional~\cite{Perdew:1996} (results listed in table~\ref{table:discontinuities}).

\section{\label{sec:results}Results and discussion}

\begin{figure}[tbp]
\begin{indented}
\item\includegraphics[width=0.8\columnwidth,clip]{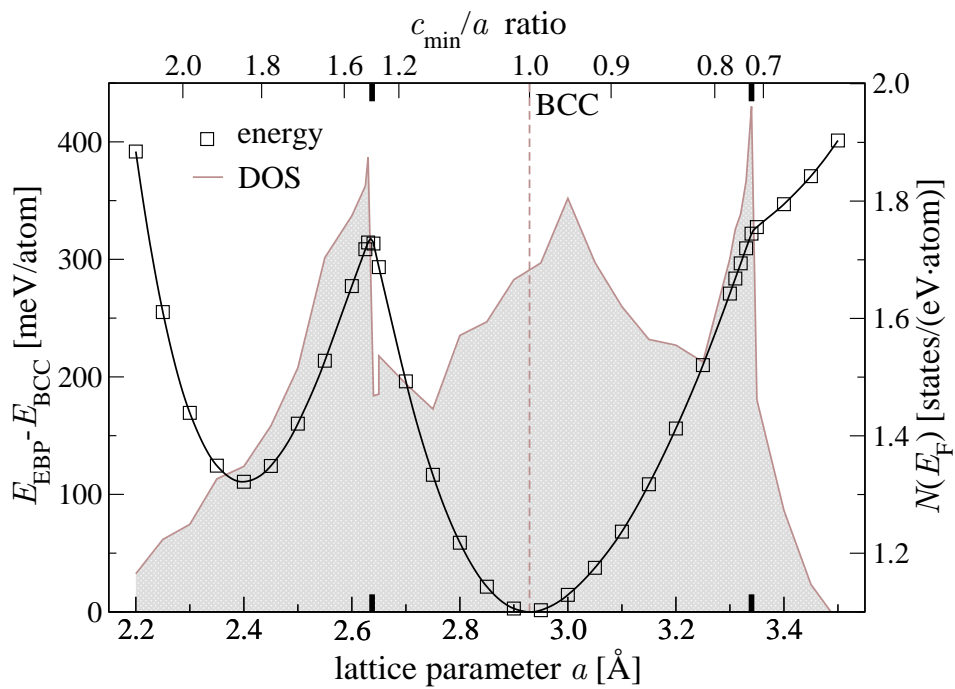}
\end{indented}
\caption{\label{fig:VEBP}
Total energy of vanadium with respect to the energy of the stable BCC phase, $E_{\rm BCC}\equiv E_0$, (left hand ordinate) and the electronic density of states (DOS) at the Fermi level, $E_{\rm F}$, (right hand ordinate) along the EBP. 
The two discontinuities of the EBP are marked by solid bars on both the $a$ and the $c_{\rm min}/a$ scale. The first discontinuity at $a_{{\rm crit},1}=2.637(3)\,\textrm{\AA}$ results in a pronounced drop in $N(E_{\rm F})$.
The second discontinuity at $a_{{\rm crit},2}=3.345(5)\,\textrm{\AA}$ is related to a peak in the DOS.
The energies are spline interpolated to guide the eye. 
The PW92 equilibrium lattice parameters of the cubic structures are
$a_{\rm BCC}=2.929\,\textrm{\AA}$; $a_{\rm FCC}=c_{\rm FCC}/\sqrt{2}=2.643\,\textrm{\AA}$. The brown broken line denotes the BCC structure.}
\end{figure}

Vanadium is one of the elements which EBP was studied in several 
investigations~\cite{Alippi:1997,Marcus:2002,Marcus:1998a,Tian:1998}. 
This provides us with the following, as we will argue incomplete, picture. 
The EBP of vanadium includes two energy minima: one corresponds to the stable BCC phase;
the other one at $a=2.400$ \AA{}, $c_{\rm min}/a=1.84$, is a local minimum, see \fref{fig:VEBP}.
This local minimum corresponds to an unstable BCT bulk state~\cite{Marcus:2002,Schoenecker:2011}, since its energy is lowered by a shear deformation.
Tian \emph{et al.}~\cite{Tian:1998,Tian:1999} however succeeded to stabilize coherently grown vanadium-films on Cu\{001\} and Ni\{001\} which structures were determined to be strained derivatives of this BCT state. 
The face-centered cubic (FCC) structure was found at a saddle point of $E(a,c)$ and 
assigned to a maximum of $E_{\rm EBP}$~\cite{Alippi:1997, Marcus:2002, Marcus:1998a, Tian:1998}.

\begin{figure}[tbp]
\begin{indented}
\item \includegraphics[width=0.8\columnwidth,clip]{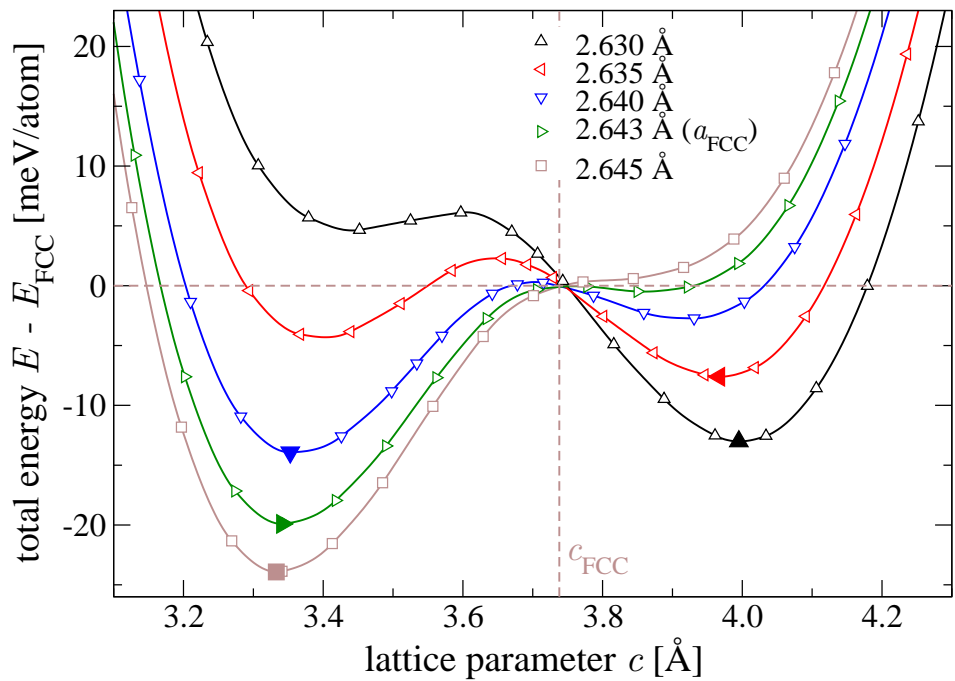}
\end{indented}
\caption{\label{fig:VEvsc}
Cuts of the total energy $E(a,c)$ of vanadium
for several fixed in-plane lattice parameters $a$ (legend) near the 
FCC saddle point.
For each $a$, the global minimum at $c_{\rm min}$ is indicated by a filled symbol.
Solid lines are spline interpolations to the data (symbols) and guide the eye.}
\end{figure}

By symmetry, the BCC structure ($c/a=1$) 
and the FCC structure ($c/a=\sqrt{2}$) are stationary points of 
$E(a,c)$~\cite{Marcus:1998a,Marcus:2002}, but the side condition 
in definition \eref{eq:mydefEBP} determines, {\em whether they are located at
the EBP or not}.
At variance with previous 
reports~\cite{Alippi:1997,Marcus:2002,Marcus:1998a,Tian:1998},
we find that the FCC structure does {\em not} belong to the EBP of vanadium, 
because it is unstable against a tetragonal distortion at fixed 
$a=a_{\rm FCC}$.
We note, that in terms of linear elasticity theory 
this instability of FCC vanadium with respect to a relaxation of $c$ corresponds to a negative elastic constant $c_{zzzz}<0$, since  $c_{zzzz}\propto \textrm{d}^2 E / \textrm{d} c^2$.
This is obvious from
the plot of $E(a_{\rm FCC},c)$ in \fref{fig:VEvsc}.
There, $E(a_{\rm FCC},c)$ has a local maximum 
at $c_{\rm FCC} = \sqrt{2}\, a_{\rm FCC}$
and the global minimum is situated at 
$c_{\rm min}=3.35\,\textrm{\AA}$, $c_{\rm min}/a=1.27$.

The characteristic feature of this instability is the occurrence of a 
double well in $E(a_{\rm FCC},c)$ vs.$\!$ $c$ with the global
minimum at $c_{\rm min}$ and a local minimum.
According to \fref{fig:VEvsc}, double wells also occur for other
values of $a$ in the vicinity of $a_{\rm FCC}$. The energies 
of both minima vary with $a$ such that the two minima are degenerate
at a critical value $a=a_{\rm crit}$.
All in all, the behavior shown in \fref{fig:VEvsc} is typical
for a first-order phase transition. One can readily read off the
energy barrier of somewhat less than 10 meV per atom, an upper
limit for the hysteresis of $\Delta a_{\rm hyst} \approx 0.015$ \AA{}, and
the discontinuity of $c_{\rm min}$, $\Delta c_{\rm min} = |c^1_{\rm min}(a_{\rm crit}) - c^2_{\rm min}(a_{\rm crit})|\approx 0.58$ \AA{}.

In general, a saddle point in $E(a,c)$ 
(here, at $(a_{\rm FCC},\sqrt{2}\, a_{\rm FCC})$) and
a discontinuity in the EBP do not have the same $a$-coordinates.
In the case of vanadium, the already described (first) discontinuity occurs 
at $2.635\,\textrm{\AA} < a_{{ \rm crit},1} < 2.640\,\textrm{\AA}<a_{\rm FCC}$.
The associated change of $c_{\rm min}/a$ amounts to $\Delta c_{{\rm min},1}/a_{{\rm crit},1} \approx 0.2$. 
Surprisingly, the EBP of vanadium shows yet another discontinuity at
$3.34\,\textrm{\AA} < a_{{\rm crit},2} < 3.35\,\textrm{\AA}$ with
$\Delta c_{{\rm min},2}/a_{{\rm crit},2}\approx 0.05$.

\begin{figure}[tbp]
\begin{indented}
\item\includegraphics[width=0.8\columnwidth,clip]{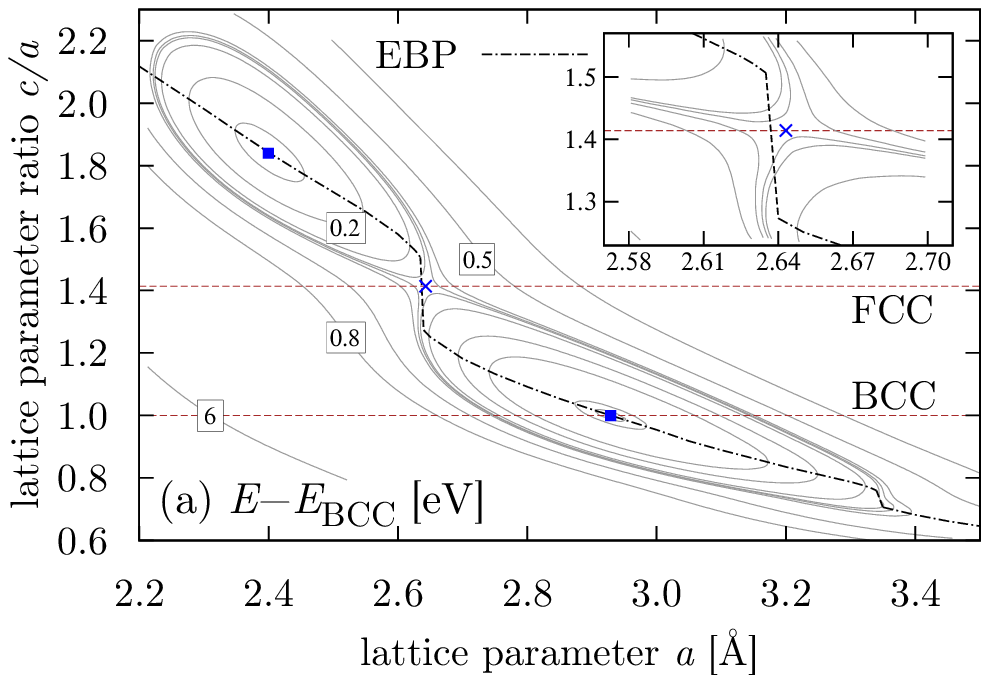}
\item\includegraphics[width=0.8\columnwidth,clip]{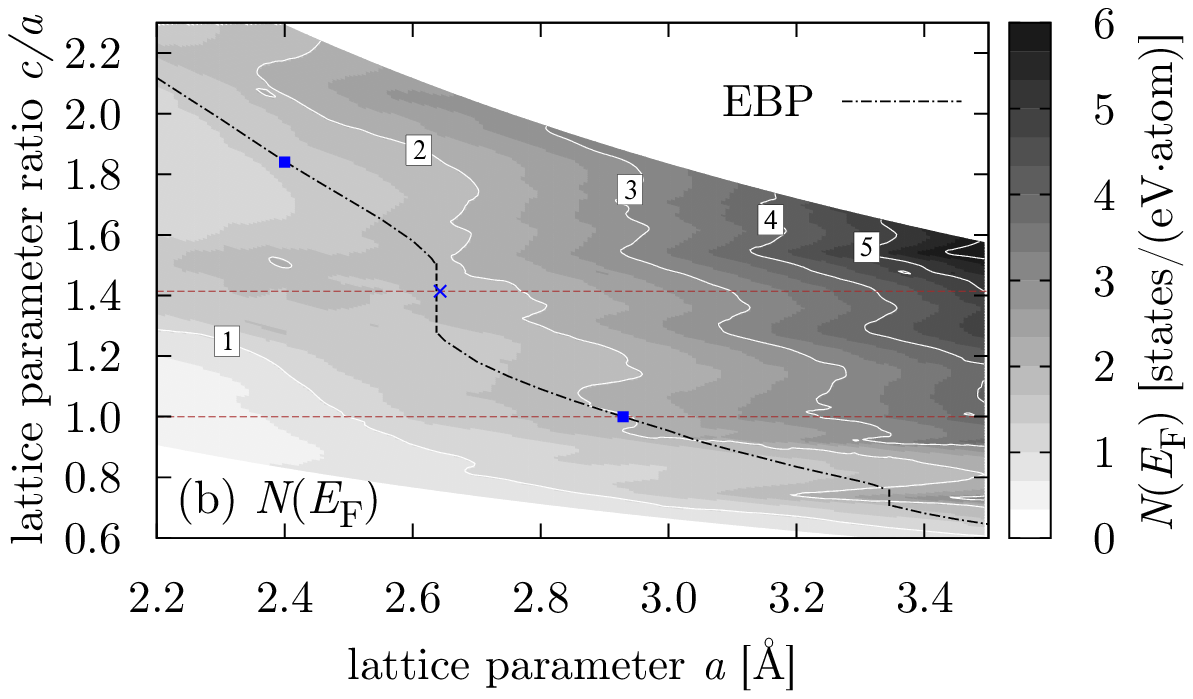}
\end{indented}
\caption{\label{fig:Vphasespace}
Contour plots in the tetragonal parameter space: (a) total energy landscape
and (b) $N(E_{\rm F})$ of vanadium. 
The two minima of $E(a,c/a)$ are indicated by ${\color{blue} \blacksquare}$ and the saddle point at the FCC structure is indicated by {\color{blue} $\times$}. 
The inset shows the saddle point region in more detail.
The global minimum corresponds to the stable BCC phase, the local minimum is a BCT structure.
The EBP (black dot-dashed line) has two discontinuities, indicated by black dashed lines. 
Brown hatched lines signal axial ratios of the cubic structures.
The energy and the density of states were calculated on a uniform $(a,c)$ grid with a distance of $0.025\,\textrm{\AA}$ between two adjacent points.}
\end{figure}

\Fref{fig:Vphasespace}a depicts a contour plot of $E(a,c/a)$ of vanadium
together with the EBP.
The two minima of $E(a,c/a)$ correspond to the two minima of $E_{\rm EBP}$
from \fref{fig:VEBP} and the FCC structure is the saddle point as 
demanded by symmetry.
The contour diagram (see inset) reveals the following: There are pairs of lines 
of constant energy, $E<E_{\rm FCC}$, where each line of the pair encloses one of the
two minima and the two lines overlap in their $a$-coordinates in the vicinity 
of the saddle point.
The double wells in \fref{fig:VEvsc} arise from this overlap.
A similar peculiarity of the energy landscape at $a \sim 3.35$ \AA{}
causes the second discontinuity and the related
kink of $E_{\rm EBP}$ visible in \fref{fig:VEBP}.

While the specific energy landscape
in the vicinity of the saddle point leads to a discontinuous EBP,
the mere presence of a saddle point --- be it symmetry
dictated or not --- does not automatically lead to a discontinuous EBP.
For example, if abscissa and ordinate axes in \fref{fig:Vphasespace}a were interchanged,
determining the EBP would mean to search for minima along horizontals, which would result in a continuous EBP including the FCC structure.

In order to analyze the origin of the discontinuities,
we consider the electronic density of states (DOS) at the Fermi level, 
$N(E_{\rm F})$, see \fref{fig:VEBP} and \fref{fig:Vphasespace}b.
The transition at $a_{{\rm crit},1}$ is accompanied by a pronounced 
change in $N(E_{\rm F})$ 
due to a significant change of the unit cell volume.
This change depends monotonously on $c$, see \fref{fig:Vphasespace}b.
Further analysis of the total energy shows that the discontinuity at
$a_{{\rm crit},1}$ is mainly driven by the kinetic energy contribution.
Since there is no obvious peculiarity in $N(E_{\rm F})$, the double well
has to be assigned to an integral effect of the whole DOS.
At the position of the second transition $a_{{\rm crit},2}$,
however, there is a distinct peak of 
$1.96\,\textrm{states/}(\textrm{eV} \cdot\textrm{atom})$ 
in $N(E_{\rm F})$ along the EBP (\fref{fig:VEBP}).
The same characteristic peak appears as tooth-like feature
in the large-$a$ region
at approximately $c/a=0.73$ in \fref{fig:Vphasespace}b.
The EBP of vanadium, which crosses this area,
becomes discontinuous in order to avoid the large $N(E_{\rm F})$
(Jahn-Teller behavior).
At the local maximum of the related double well of $E(a_{{\rm crit},2},c)$
(not shown), $N(E_{\rm F}) = 2.4\,\textrm{states/}(\textrm{eV}\cdot \textrm{atom})$.


\begin{table}[htb]
\caption{\label{table:discontinuities}List of elements with discontinuous
EBPs in PW92, which were cross-checked in PBE96. The exchange-correlation functional (XC), the position of the discontinuity, $a_{\rm crit}$,
the related out-of-plane lattice parameters $c^{1/2}_{\rm min}$,
the change of $c_{\rm min}$ across the discontinuity, $\Delta c_{\rm min}$, the relative volume
change $\Delta V/V=2\Delta c_{\rm min}/(c^{1}_{\rm min}+c^{2}_{\rm min})$, and the strain energy at $a_{\rm crit}$ with respect to the energy of the ground state are tabulated.}
\begin{indented}
\item[]\begin{tabular}{cr*{6}{c}}
\br
 element & XC & $a_{\rm crit}$  & $c^{1}_{\rm min}$ & $c^{2}_{\rm min}$ & $\Delta c_{\rm min}$  & $\Delta V/V$ & $E_{\rm EBP}(a_{\rm crit}) - E_0$  \\
   & & [$\textrm{\AA}$] & [$\textrm{\AA}$] & [$\textrm{\AA}$] & [$\textrm{\AA}$] & \% &[meV/atom] \\
\mr
   V  & PW92 & 2.637(3) & 3.97 & 3.39 & 0.58 & 16 & 313  \\
     & PBE96  & 2.698(2) & 3.96 & 3.51 & 0.44 &  12 & 280  \\
   V  & PW92 & 3.345(5) & 2.54 & 2.37 & 0.17 & 7 & 323  \\
     & PBE96  & 3.447(2) & 2.59 & 2.46 & 0.13 & 5 &  330 \\
   Nb & PW92 & 2.933(2) & 4.38 & 3.76 & 0.62 & 15 & 390 \\
    & PBE96  & 2.988(2) & 4.41 & 3.87 & 0.54 & 13 & 354 \\
   Ru & PW92 & 3.010(1) & 3.26 & 3.03 & 0.23 & 7 & 685 \\
    & PBE96  & 3.062(2) & 3.31 & 3.05 & 0.26 & 8 & 627 \\
   La & PW92 & 4.043(1) & 4.50 & 3.67 & 0.83 & 20 & 178 \\
    & PBE96  & 4.232(2) & 4.65 & 3.91 & 0.74 & 17 &  103 \\
   Os & PW92 & 3.063(1) & 3.28 & 3.02 & 0.26 & 8 & 907 \\
    & PBE96  & 3.104(1) & 3.39 & 3.03 & 0.36 & 11 & 900 \\
   Ir & PW92 & 3.077(2) & 3.23 & 2.89 & 0.24 & 11 & 630 \\
    & PBE96  & 3.118(1) & 3.31 & 2.93 & 0.38 & 12 & 818 \\
\br
\end{tabular}
\end{indented}
\end{table}

We searched the EBPs of the 30 elements listed above
for the occurrence of first-order transitions in the same way 
as described for the case of vanadium~\cite{Schoenecker:2011}.
The search space was restricted to $0.7 < c/a < 2.1$ and had a
resolution $\Delta a = 0.05$\,\AA{} ($\Delta a$ was subsequently reduced to pinpoint $a_{\rm crit}$).
Seven transitions were found in this way, see table~\ref{table:discontinuities}.
Apart from vanadium, also the group-5 element niobium (but not tantalum) and
the elements ruthenium, lanthanum, osmium, and iridium exhibit discontinuous EBPs.
For La and Ir, we identified a negative $c_{zzzz}$
of the BCC structure as the driving force of the instability,
while in Nb the FCC structure has $c_{zzzz} < 0$.
BCC Ru and Os possess small positive $c_{zzzz}$ but the discontinuities of the EBP occur in the vicinity of the BCC structure at a lattice parameter being slightly smaller and larger than $a_{\rm BCC}$, respectively. 
No second instability was found for any element except vanadium.
All seven first-order transitions were also found in additional calculations relying on PBE96, see table~\ref{table:discontinuities}. Compared to the PW92 results, $a_{\rm crit}$ is shifted to larger values in PBE96.

The present work focuses on non-magnetic elements, where
we found first-order martensitic transitions under epitaxial strain
due to either a large peak in $N(E_{\rm F})$
or a more subtle interplay of all contributions to the total energy
in the vicinity of a saddle point.
In order to find out whether the epitaxial relation of a particular substance
is continuous or not, one usually has to calculate this relation in detail.
An FCC or BCC structure with negative $c_{zzzz}$ always results in a discontinuous EBP but a (small) positive $c_{zzzz}$ does not
exclude a discontinuity.

In the case of magnetic systems,
high-spin to low-spin transitions~\cite{Rosner:2006}
or a change of the ground state magnetic order~\cite{Qiu:2001} are usually
accompanied by discontinuities of the lattice geometry.
An example might be the $a$-dependence of the magnetic ground state of BCT Fe
found by Qiu {\em et al.}~\cite{Qiu:2001}.
These authors did however not discuss the consequence of a transition into
a different magnetic state for the lattice parameter $c$ of coherent films.

We now return to the case of vanadium to demonstrate, how strongly
the structural discontinuity may alter another particular intrinsic
property, namely the superconducting critical temperature $T_{\rm c}$.
Vanadium is a well-known superconductor and
possesses one of the highest $T_{\rm c}$-values among all elements,
$T_{\rm c} = 5.4$\,K at normal pressure~\cite{Buzea:2005}.  
A significant change of $T_{\rm c}$ may be expected at $a_{{\rm crit},1}$,
since $N(E_{\rm F})$ jumps by 
$0.2\,\,\textrm{states/}(\textrm{eV}\cdot\textrm{atom})$, see \fref{fig:VEBP}.

The calculations for $T_{\rm c}$ of BCT vanadium were carried out with the 
ABINIT-DFT package in the version 6.12~\cite{Gonze:2005,Gonze:2009} 
employing a pseudopotential (PP) generated with the FHI98PP 
code~\cite{Fuchs:1999} for PW92.
With this PP, we found slightly shifted values of 
$a_{\rm BCC}=2.963\,\textrm{\AA}$ and
of $a_{{\rm crit},1}=2.623\,\textrm{\AA}$ compared with the FPLO results.
The phonon properties were obtained through density-functional perturbation theory~\cite{Baroni:2001} as implemented in ABINIT~\cite{Gonze:1997,Gonze:1997b}.
To compute the superconducting properties of V, we used a plane-wave kinetic-energy cutoff of 650\,eV and a $52 \times 52 \times 52$ mesh in the full Brillouin zone for linear tetrahedron integrations. The phonon quantities are sampled by eight wave vectors in the full Brillouin zone. 

We verified the accuracy of our approach for BCC vanadium for which we obtained an electron-phonon coupling constant $\lambda=0.95$ and a logarithmic-averaged phonon frequency $\omega_{\rm log}=288\,\textrm{K}$. For comparison, tunneling experiments found $\lambda=0.82$~\cite{Wolf:1985}.
We derived a reasonable value of $T_{\rm c}= 3.2$\,K for BCC vanadium according to the Allen-Dynes modified McMillan formula~\cite{Allen:1975}. This equation contains one parameter, the effective Coulomb repulsion $\mu^\ast$, which we set to a value $\mu^\ast = 0.13$ commonly used for transition metals~\cite{Delaire:2008}.

$T_{\rm c}$ across the discontinuity $a_{{\rm crit},1}$ was computed for the points ($(a,c_{\rm min})$ coordinates) ($2.595\,\textrm{\AA},4.128\,\textrm{\AA}$) and ($2.646\,\textrm{\AA},3.545\,\textrm{\AA}$).
$\lambda$ changes by $\Delta \lambda=0.20$ and $T_{\rm c}$ by $\Delta T_{\rm c}=0.6\,\textrm{K}$ or 20\,\% across $a_{{\rm crit},1}$
(from 3.5\,K at $a=2.595\,\textrm{\AA}$ to 4.1\,K at $a=2.646\,\textrm{\AA}$). The actual value of $T_{\rm c}$ may depend on the film thickness as
this was found in experiments for BCC vanadium 
films~\cite{Teplov:1976,Teplov:1981}, but we expect that the relative change will not substantially be affected by this dependence.

\section{Conclusions}

We have shown that coherently grown epitaxial films can exhibit
a discontinuous relation between in-plane and out-of-plane lattice parameters
in a strain-driven first-order phase transition. 
In particular, we predict that BCT bulk-like films of the elements 
V, Nb, Ru, La, Os, and Ir undergo martensitic lattice transformations induced 
by epitaxial strain with related volume changes up to 20\%. 
As a consequence, the intrinsic film properties can be switched, if one succeeds
to control the substrate lattice parameter appropriately.
We demonstrated this possibility by the example of vanadium, where the
superconducting critical temperature changes by about 20\% at a
critical strain value.

Since the mechanisms of the discussed instabilities may be present in
arbitrary metallic systems, our results are not limited to monoatomic overlayers,
BCT films or specific ground states. 
Verification of the predicted
transitions should be possible by advanced diffraction methods applied
to coherently grown overlayers on a series of substrates with
quasi-continuous lattice spacings.
An indirect verification could be achieved by measuring any other 
structure-dependent physical property of the film material.

\ack

We thank K D\"orr, S Gemming, S Haindl, S F\"ahler, and A M\"obius for discussions.

\section*{References} 


\providecommand{\newblock}{}

\end{document}